\documentclass[aps,prl,twocolumn,showpacs,superscriptaddress,floatfix]{revtex4}
\usepackage{graphicx} 
\def\be{\begin{equation}} 
\def\ee{\end{equation}} 
\def\b8{$^8$B} 
\def\pb208{$^{208}$Pb} 
\def\b7{$^7$Be} 
\def\s17{$S_{17}$} 
\begin{document} 
  
\title{Charge radius and dipole response of $^{11}$Li}
\author{H. Esbensen}
\affiliation{Physics Division, Argonne National Laboratory, Argonne, Illinois 60439, USA}
\author{K. Hagino}
\affiliation{Department of Physics, Tohoku University, Sendai, 980-8578,
Japan}
\author{P. Mueller}
\affiliation{Physics Division, Argonne National Laboratory, Argonne, Illinois 60439, USA}
\author{H. Sagawa}
\affiliation{Center for Mathematical Sciences, University of Aizu,
Aizu-Wakamatsu, Fukushima 965-8560, Japan}
\date{\today} 
  
\begin{abstract} 
We investigate the consistency of the measured charge radius and dipole 
response of $^{11}$Li within a three-body model. We show how these 
observables are related to the mean square distance between the $^9$Li 
core and the center of mass of the two valence neutrons. 
In this representation we find by considering the effect of smaller
corrections that the discrepancy between the results of the two 
measurements is of the order of 1.5$\sigma$. 
We also investigate the sensitivity to the three-body structure
of $^{11}$Li and find that the charge radius measurement favors a model 
with a 50\% s-wave component in the ground state of the two-neutron halo, 
whereas the dipole response is consistent with a smaller s-wave component 
of about 25\% value.
\end{abstract} 
 
\pacs{21.10.Ft,21.45.+v, 21.60.Gx,25.60.-t} 
\maketitle 

\section{Introduction}

The properties of the two-neutron halo nucleus $^{11}$Li have 
been discussed in numerous theoretical and experimental papers 
but knowledge about its structure is still uncertain. 
In the past year, the results of two important measurements have 
been published, namely, the RMS (root-mean-square) charge radius 
of $^{11}$Li obtained from laser spectroscopy \cite{sanchez} and 
the dipole response, which was probed by Coulomb dissociation 
on a Pb target \cite{naka}.
The purpose of this paper is to investigate whether the two 
measurements can be explained simultaneously within a three-body 
model and to see what are the implications of the two results for 
the structure of $^{11}$Li.

The nucleus $^{11}$Li is an excellent example of a so-called 
borromean system, a bound three-body system in which none of 
the two-body subsystems form a bound state.
Thus $^{11}$Li can be viewed as a three-body system consisting 
of a $^9$Li core and two valence neutrons where neither 
$^{10}$Li nor the dineutron system has a bound state.  
The nucleus $^{11}$Li has only one bound state with a two-neutron 
separation energy of about 300 keV.

Theoretical studies of $^{11}$Li have primarily been based 
on three-body models of the two valence neutrons interacting
with the $^9$Li core, and it has been a major challenge over 
the past 10--15 years to obtain 
information about the neutron-core interaction, i.~e., about 
the scattering states in the unbound nucleus $^{10}$Li.
Early studies \cite{ann91} assumed a dominant p-wave structure 
of the two-neutron halo, based on a rather high-lying p-wave 
resonance in $^{10}$Li. 
Another model \cite{zhukov} assumed a shallow neutron-core 
potential, which does not have any bound states, and this
resulted in a strongly s-wave dominated ground state of the 
two-neutron halo.
In order to explore the structure of $^{11}$Li, a wider range
of models were developed \cite{thompson} and compared to 
measurements.

A better calibration of three-body models for $^{11}$Li became 
possible with an accurate measurement of the two-neutron separation
energy, $S_{2n}$ = 295$\pm$15 keV \cite{young1},
and a production measurement \cite{young2} which probed the continuum 
of $^{10}$Li. The latter measurement suggested a p-wave resonance at 
about 540 keV and some influence of s-wave scattering near threshold.
Let us also mention that the quadrupole moments of $^9$Li and $^{11}$Li 
are the same within the 15\% experimental uncertainty \cite{neugart};
this can be taken as a justification for using three-body models.

An analysis of the $\beta$-decay of $^{11}$Li \cite{otsuka} showed 
that about 45--55 \% of the two-neutron halo must be in $p_{1/2}$ 
orbits, whereas the remaining part would most likely occupy s-waves.
Analyses of the momentum distributions produced in high-energy
breakup reactions also suggested a large s-wave component, 
from 20--40 \% \cite{prc57} to 35-55 \% \cite{simon}. 
In fact, there seems to be a consensus toward a large s-wave 
component in two-neutron ground state, a component that is much
larger than what was expected.
This feature may be  related to the famous parity inversion in the 
neighboring nucleus $^{11}$Be, where the ground state is a $1/2^+$
state and not a $1/2^-$ state as one naively would expect for a 
p-shell nucleus. It is of interest to see how the recent 
charge radius \cite{sanchez} and dipole response \cite{naka}
measurements fit into this trend.

\section{Three-body model interpretation}

There is a very close relationship within a three-body model between 
the charge radius and the dipole response of a two-neutron halo
since they are both probes of the distance between the $^9$Li core
and the center-of-mass of the dineutron system.
The mean square charge radius $\langle r_p^2(Z,A) \rangle$ 
for point-nucleons, for example, can be expressed in terms of the 
charge radius of the core nucleus as follows,
\begin{equation}
\langle r_p^2(Z,A)\rangle = \langle r_p^2(Z,A-2) \rangle 
+ \Bigl(\frac{2}{A}\Bigr)^2 \  \langle r_{c,2n}^2 \rangle,
\label{ppchr}
\end{equation}
where the second term is the correction which is caused by the 
center of mass motion of the core nucleus in the presence of the 
two valence neutrons. The correction is proportional to the mean 
square distance, $\langle r_{c,2n}^2\rangle$, between the core
and the center of mass of the dineutron system.

The total strength of the dipole response of a two-neutron halo nucleus 
is approximately given by the cluster sum rule \cite{ann91},
\begin{equation}
B(E1) = \frac{3}{4\pi} \ \Bigl(\frac{Ze}{A}\Bigr)^2 \
4\ \langle r_{c,2n}^2\rangle,
\label{csr}
\end{equation}
which is also expressed in terms of the mean square distance,
$\langle r_{c,2n}^2\rangle$, between the core and the dineutron system.
The sum rule assumes that the total dipole strength can be
calculated by closure, which includes dipole transitions
to the Pauli blocked core states. The effect of Pauli blocking is a 
minor but not insignificant correction, as we discuss in the next section. 
However, if we ignore it, we see that the charge radius and dipole 
response measurements are closely related, since they are both probes 
of the core-dineutron distance. 

\begin{figure}
\includegraphics [width = 8.5 cm]{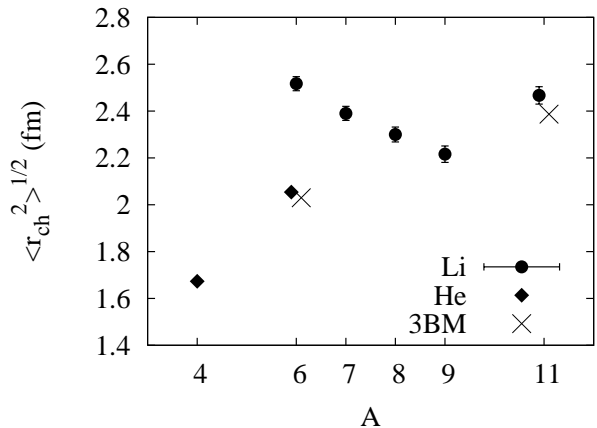}
\caption{Measured charge radii of He \cite{wang} and Li 
\cite{sanchez,ewald} isotopes are compared to the predictions 
of the three-body models (3BM) of $^6$He and $^{11}$Li 
discussed in Ref. \cite{prc56}.}
\label{chrad}
\end{figure} 

To make contact with the measured charge radius $r_{ch}$ we
must correct the point-proton charge radii of Eq. (\ref{ppchr}) 
for the finite sizes of protons and neutrons.
We should also consider the so-called Darwin-Foldy contribution
and the effect of the spin-orbit charge density discussed in 
Ref. \cite{friar}. We therefore have the expression,
\begin{equation}
\langle r_{ch}^2 \rangle = \langle r_p^2 \rangle 
+ \langle R_p^2\rangle + \frac{A-Z}{Z} \ \langle R_n^2 \rangle
+ \frac{3\hbar^2}{4(mc)^2} + 
\langle r^2\rangle_{\rm so},
\label{sizecor}
\end{equation}
where $\langle R_p^2\rangle$ = 0.757(14) fm$^2$ and
$\langle R_n^2\rangle$ = -0.1161(22) fm$^2$ are the
mean square charge radii of protons and neutrons \cite{PLB592},
respectively, and $3\hbar^2/[4(mc)^2]$ is the Darwin-Foldy term.

Inserting Eq. (\ref{ppchr}) into Eq. (\ref{sizecor}) we obtain
the following expression for the difference between the mean square 
charge radii of the two-neutron halo nucleus and the core nucleus,
$$
\delta \langle r_{ch}^2 \rangle =
\langle r_{ch}^2(Z,A) \rangle - \langle r_{ch}^2(Z,A-2) \rangle 
$$
\begin{equation}
= \Bigl(\frac{2}{A}\Bigr)^2 \ \langle r_{c,2n}^2 \rangle 
- \frac{0.232}{Z}
+ \langle r^2\rangle_{\rm 2n}^{\rm so}.
\label{chreq}
\end{equation}
It is seen that the proton charge radius $R_p$ in Eq. (\ref{sizecor}) 
drops out of Eq. (\ref{chreq}), and so does the constant Darwin-Foldy 
term. The only two corrections that survive,
$-0.232/Z$ and $\langle r^2\rangle_{2n}^{\rm so}$,
are due to the non-zero, mean square charge radius of a neutron 
and the spin-orbit charge density of the two valence neutrons.

In the following we ignore the spin-orbit correction to the charge 
radius except when otherwise explicitly stated. The reason is that this 
correction is model dependent, so it is not obvious how one should convert 
the measured isotope shift into a mean square, core-dineutron distance. 
One can calculate the spin-orbit correction in different models
from the explicit expressions that are given in Ref. \cite{friar}.
In the shell model for spherical nuclei, with two valence neutrons 
occupying an unfilled $(l,j)$ sub-shell, the spin-orbit correction
is 
\begin{equation}
\langle r^2 \rangle_{2n}^{\rm so} =
\frac{2\mu_n}{Z} \ 
\Bigl(\frac{\hbar}{m}\Bigr)^2
\langle {\bf l\cdot s}\rangle, 
\label{soc}
\end{equation}
where $\mu_n$ = -1.913 in the neutron magnetic moment, 
$m$ is the neutron mass, and 
$$
\langle {\bf l\cdot s}\rangle =j(j+1)-l(l+1)-3/4.
$$
In three-body models one would  
have to calculate $\langle {\bf l\cdot s}\rangle$ numerically as 
an average value, since the $0^+$ ground state of the two-neutron 
halo contains many $(l,j)$ single-particle components \cite{prc56}.

\begin{table}
\caption{The measured charge radius of $^6$He \cite{wang},
the change in the mean square charge radius of $^{11}$Li and $^9$Li,  
and the mean square distance between the $^4$He core and the 
two-neutron halo in $^6$He are compared to results of three-body 
models (3BM) and GFMC calculations.
The spin-orbit correction, Eq. (\ref{soc}), was ignored
and the charge radius of  $^4$He was set to 1.673(1) fm.}
\begin{ruledtabular}
\begin{tabular} {|c|c|c|c|}
Nucleus & $\langle r_{ch}^2\rangle^{1/2}$ (fm) & 
$\delta \langle r_{ch}^2\rangle$ (fm$^2$) &
$\langle r_{c,2n}^2\rangle$ (fm$^2$) \\
\colrule
 $^6$He exp. \cite{wang} & 2.054(14) & 1.42(5)  & 13.8 $\pm$ 0.5 \\
 3BM \cite{prc56}        & 2.036     & 1.35     & 13.2   \\ 
 3BM \cite{cobis}        & 2.011     & 1.25     & 12.3    \\ 
GFMC \cite{pieper}       & 2.08(4)   & 1.49(15) & 14.5 $\pm$ 1.5 \\ 
\end{tabular}
\end{ruledtabular}
\end{table}

We show in Fig. \ref{chrad} the measured RMS charge radii for the 
helium \cite{wang} and lithium \cite{ewald,sanchez} isotopes. 
The results for $^6$He and $^{11}$Li are compared with the three-body
model calculations of Ref. \cite{prc56}. These calculations employed 
a density-dependent contact interaction to simulate low-energy
$nn$ scattering,
and the $^4$He-neutron Hamiltonian was calibrated to reproduce the 
known low-energy neutron-$\alpha$ scattering phase shifts. 
The calculations for $^{11}$Li are discussed below.

The measured charge radius of $^6$He and the mean-square, 
core-dineutron distance we obtain when we ignore the spin-orbit 
correction in Eq. (\ref{chreq}) are compared in Table I
to the results of different models, namely, 
two three-body models of Refs. \cite{prc56,cobis}, and a 
recent GFMC (Greens function Monte Carlo) calculation \cite{pieper}.
The latter is an improvement over the results that were
published in Ref. \cite{pudle}. 

Let us estimate the spin-orbit correction to the $^6$He-$^4$He 
mean-square charge radius difference in the extreme limit where 
the valence neutrons occupy a pure $(p_{3/2})^2$ configuration. 
Then $\langle {\bf l\cdot s}\rangle$ = 1, and we obtain 
from Eq. (\ref{soc}) the correction
$\langle r^2\rangle_{2n}^{\rm so}$ = -0.085 fm$^2$. 
Another estimate is to evaluate the average spin-orbit correction 
in the three-body model developed in Ref. \cite{prc56}, since we 
know in this case the occupation probabilities of the 
single-particle orbits. 
The model quoted in line 5, Table II of Ref. \cite{prc56}, has 83\% 
of the two-neutron halo in $(p_{3/2})^2$ orbits. Considering all 
orbits of the halo we obtain the average value 
$\langle {\bf l\cdot s} \rangle$ = 0.82.
This implies the spin-orbit correction 
$\langle r^2\rangle_{2n}^{\rm so}$ = -0.07 fm$^2$,
which would bring the GFMC calculation into perfect agreement 
with the measurement, whereas the three-body model \cite{prc56} 
would be off by 10\%, which is a discrepancy of almost $3\sigma$.

The discrepancy between the measured and calculated charge radius 
of $^{11}$Li which can be seen in Fig. \ref{chrad} may reflect 
uncertainties in the neutron-core Hamiltonian that we have used. 
The neutron halo ground state contains in this case 23\% 
s-waves and is therefore referred to as the $s23$ model below.
The model Hamiltonian \cite{prc56} was calibrated to reproduce the 
measured two-neutron separation energy \cite{young1} and also the 
p-wave resonance structure observed in Ref.  \cite{young2}, but 
there is still some uncertainty in the s-wave strength which we 
explore below.

\section{Dipole response}

\begin{figure}
\includegraphics [width = 8 cm]{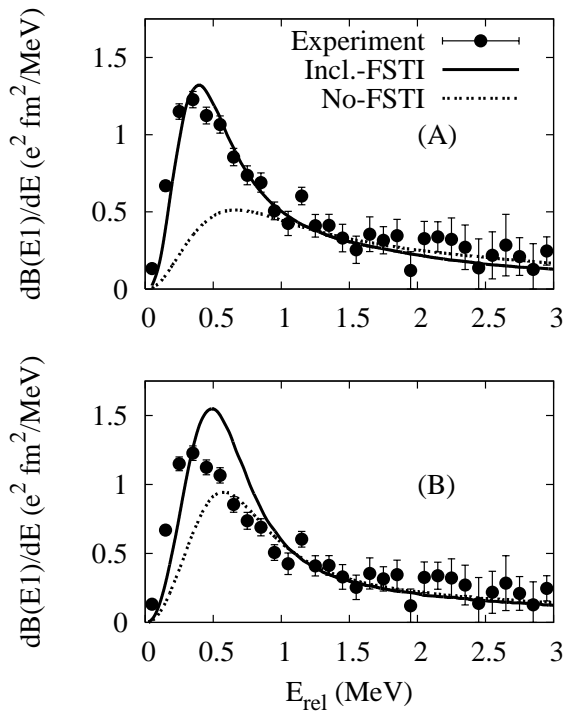}
\caption{ 
The measured dipole response of $^{11}$Li \cite{naka} is compared 
(A) to the calculations of Ref. \cite{npa542}, with (solid) 
and without (dashed) the effect of the final state $nn$ interaction,
and (B) to calculations that are based on the $s23$ model described 
in the text.} 
\label{dipole}
\end{figure}

The measured dipole strength distribution \cite{naka} is compared 
in Fig. \ref{dipole}A to the prediction of an old three-body model 
of $^{11}$Li \cite{npa542}. The calculated distributions 
include the effect of the experimental energy resolution \cite{naka}.  
Although this model has a rather small two-neutron separation 
energy of 200 keV and did not include the recoil effects in a 
three-body system, it was able to reproduce fragmentation data 
at 800 MeV/nucleon fairly well \cite{prc46},
and Fig. \ref{dipole}A shows that it also produces a dipole response 
that is in surprisingly good agreement with the measurement \cite{naka}. 
The strong peak near 300--400 keV (solid curve) is produced by the strong 
attractive interaction between the neutrons in the final state. 
If this final state interaction is set to zero
we obtain the dashed curve. 

The measured dipole strength distribution \cite{naka} is compared
in Fig. \ref{dipole}B to new calculations that are based on the 
$s23$ model \cite{prc56}. The calculations (with and without the
effect of the final state $nn$ interaction) include the experimental 
energy resolution as done in Ref.  \cite{naka}.
The ground state of the $s23$ model has a realistic two-neutron 
separation of 295 keV and includes recoil effects in the 
three-body system exactly, c.~f. the last term of Eq. (3.1)
in Ref. \cite{prc56}.  The recoil effects are treated 
approximately in the three-body final state of the dipole 
response by ignoring the off-diagonal component 
$\vec{p}_1\cdot\vec{p}_2/(A_cm)$ (the last term in Eq. (3.3) 
of Ref. \cite{prc56}), whereas the diagonal term 
$\vec{p}_1^2/(2A_cm) +\vec{p}_2^2/(2A_cm)$ is included
through the reduced mass. 
With this approximation, the continuum dipole response can be 
computed with the method of Ref. \cite{npa542}. 
We have checked the accuracy of this approximation with the 
discretized dipole strength function of Ref. \cite{hagprc72}
and have confirmed that the approximation 
works well for the $^{11}$Li nucleus. 
The calculated peak (solid curve in Fig. \ref{dipole}B) is higher 
and shifted slightly toward higher energies in comparison to the 
data but the overall strength is very reasonable.

The total dipole strength that was measured up to a 3 MeV
relative energy is $B(E1)_{\rm exp}$ = 1.42 $\pm$ 0.18 e$^2$fm$^2$.
The calculated dipole strength up to 3 MeV, 
$B(E1,E_{rel}\leq 3 \ {\rm MeV})_{\rm cal}$, is 1.26 e$^2$fm$^2$
in the old model (Fig. \ref{dipole}A) and 1.38 e$^2$fm$^2$ in the 
new $s23$ model (Fig. \ref{dipole}B). 
Since the measured and calculated dipole strengths shown in Fig.
\ref{dipole} do not differ much,
it seems reasonable to estimate the mean-square, core-dineutron
distance associated with the experiment by the simple scaling,
\begin{equation}
\langle r_{c,2n}^2\rangle \approx
\frac{B(E1, E_{rel}\leq 3 \ {\rm MeV})_{\rm exp}}
{B(E1, E_{rel}\leq 3 \ {\rm MeV})_{\rm cal}}
\ \langle r_{c,2n}^2\rangle_{\rm 3BM},
\label{scaling}
\end{equation}
in terms of the mean-square distance $\langle r_{c,2n}^2 \rangle_{3BM}$
we obtain in the three-body model. 
We note that this scaling method is consistent with the cluster sum rule
but it does not necessarily require that the total dipole strength of 
the model is given by the cluster sum rule. 
We emphasize that the calculated dipole strength, which we insert into 
Eq. (\ref{scaling}), is calculated in a model that respects the 
Pauli principle, whereas the cluster sum rule does not.

The scaling method (\ref{scaling}) gives essentially the same result 
independent of which of the two models we use (the old model or the new 
$s23$ model). The average value is shown in the last line of Table II
and it represents the mean-square, core-dineutron distance we extract 
from the measured dipole strength distribution.
The last line of Table II also gives the difference in the mean-square 
charge radius of $^{11}$Li and $^9$Li, which we derive from 
Eq. (\ref{chreq}) by ignoring the spin-orbit correction.

\begin{table}
\caption{The difference between the measured mean square charge radii 
of $^{11}$Li and $^9$Li \cite{sanchez}, 
and the mean square distance between the $^9$Li core and the 
two-neutron halo (extracted from Eq. (\ref{chreq}) for
$\langle r^2\rangle_{2n}^{\rm so}$=0), are compared to the results 
of three-body models (3BM), and the values extracted from the 
Coulomb dissociation (CD) experiment \cite{naka}.
The assumed charge radius  of $^9$Li was 2.216(35) fm.
The last column shows the mean square distance between the two 
valence neutrons.}
\begin{ruledtabular}
\begin{tabular} {|c|c|c|c|}
Nucleus & $\delta\langle r_{ch}^2\rangle$ (fm$^2$) & 
$\langle r_{c,2n}^2\rangle$ (fm$^2$) &
$\langle r_{n,n}^2\rangle$ (fm$^2$) \\
\colrule
$^{11}$Li exp \cite{sanchez} & 1.175(124) & 37.9 $\pm$ 3.7 & \\
revised   \cite{moro}        & 1.104(85)  & 35.7 $\pm$ 2.6 & \\
\colrule
Old 3BM \cite{ann91}         & 0.728    & 24.35 & 39.0 \\
s05 3BM \cite{prc56}         & 0.541    & 18.7  & 42.8 \\
s23 3BM \cite{prc56}         & 0.789    & 26.2  & 45.9 \\
s32 3BM new                  & 0.895    & 29.4  & 51.6 \\
s50 3BM \cite{prc57}         & 1.120    & 36.2  & 70.1 \\
\colrule
CD exp. \cite{naka}          & 0.82(11) & 27.2 $\pm$ 3.5 & \\
\end{tabular}
\end{ruledtabular}
\end{table}

We note that the RMS core-dineutron distance quoted in 
Ref. \cite{naka}, which is 5.01 $\pm$ 0.32 fm, 
is smaller than the 5.22 fm value we obtain from the last 
line of Table II. The smaller size is the result of identifying
the estimated total dipole strength (1.78 e$^2$fm$^2$) with the 
cluster sum rule, Eq. (\ref{csr}).
An even smaller size was obtained in Ref. \cite{moro} by fitting 
the measured dipole strength distribution of Ref. \cite{naka},
and identifying the total strength with the cluster rule. 
The result (Eq. (20) of Ref. \cite{moro}) translates into an RMS 
core-dineutron distance of 4.73 fm.

From an experimental point of view, extreme care must be taken when 
determining the total dipole strength associated with excitations
of the halo, and when translating this strength into the size of the halo. 
This cannot be done accurately without some theoretical guidance
because the dipole strength can only be resolved at low excitation 
energies. Moreover, the cluster sum rule (\ref{csr}), which is 
sometimes used to determine the size of the halo, is not exact
because it ignores the Pauli blocking of some of the final states
as we discussed earlier.
In the old three-body model of Refs. \cite{ann91,npa542}, for example, 
the total strength obtained by numerically integrating the calculated
dipole strength distribution is 1.57 e$^2$fm$^2$, whereas the cluster 
sum rule strength is 1.73 e$^2$fm$^2$.
That implies that the total strength is reduced by 10\% compared to 
the cluster sum rule because of Pauli blocking.

\section{Charge radius measurement}

The measured charge radius of $^{11}$Li is 2.467(37) fm \cite{sanchez}.
We emphasize that the uncertainty in the measured charge radius is 
partly due to the absolute calibration of one of the isotopes ($^7$Li). 
The difference between the mean square charge radii for different isotopes 
is therefore much more accurately determined.
This is a great advantage for our discussion of the halo because the 
mean square distance between the $^9$Li core and the dineutron is directly 
related, according to Eq. (\ref{chreq}), to the difference 
$\delta\langle r_{ch}^2\rangle$ between the 
mean square charge radii of $^{11}$Li and $^9$Li.
We have considered this feature in our determination of the 
uncertainties on the values of $\delta\langle r_{ch}^2\rangle$ and
$\langle r_{c,2n}^2\rangle$ shown in Table II.

The change in the mean square charge radius from the reference
nucleus $^7$Li was obtained from the measured isotope shift
$\delta \nu_{\rm IS}^{\rm exp}(A,7)$
and the calculated so-called finite mass correction
$\delta \nu_{\rm IS}^{\rm MS}(A,7)$
according to the expression \cite{sanchez}
\begin{equation}
\langle r_{ch}^2(A) \rangle- \langle r_{ch}^2(7)\rangle =
\frac{\delta \nu_{\rm IS}^{\rm exp}(A,7)-\delta \nu_{\rm IS}^{\rm MS}(A,7)}
{1566.1 \ {\rm kHz}} \  {\rm fm}^2.
\end{equation}
The finite mass corrections that were used in Ref. \cite{sanchez} 
have recently been reevaluated \cite{moro}.  Combining these new 
corrections with the measured isotope shifts of Ref. \cite{sanchez} 
one obtains the `revised' charge radius and core-dineutron distance 
shown in Table II. 
The important quantity to our discussion is the size of the halo
which is here represented by $\langle r_{c,2n}^2\rangle$.
We see that the value we obtain from the Coulomb 
dissociation experiment is smaller than the values we obtain from 
the two interpretations \cite{sanchez,moro} of the charge radius 
measurement. The deviation is in both cases a $2\sigma$ discrepancy.

In Table II we also give the results we obtain in various three-body 
models of $^{11}$Li, ranging from the `Old' three-body model of 
Ref. \cite{ann91} to the $s50$ model of Ref. \cite{prc57}, which gives 
a 50\% s-wave component in the ground state of the two-neutron halo.
It is seen that the $s50$ model is in agreement with the charge radius 
measurement \cite{sanchez},
whereas the $s23$ model of Ref. \cite{prc56} (with 23\% s-waves 
in the halo ground state) is consistent with the CD experiment.

Let us finally estimate the spin-orbit correction, Eq. (\ref{soc}),
which we have ignored so far when applying Eq. (\ref{chreq}). 
In the extreme model, where the two valence neutrons occupy the 
$(p_{1/2})^2$ configuration, the value of
$\langle {\bf l\cdot s}\rangle$ is $-2$, and from Eq. (\ref{soc}) 
we obtain  $\langle r^2\rangle_{2n}^{\rm so}$ = 0.113 fm$^2$.
In the $s23$ model, where 61\% of the halo is in the $(p_{1/2})^2$ 
configuration, 23\% are s-waves, and 16\% are in higher $(l,j)$ orbits,
we obtain the average value $\langle {\bf l\cdot s}\rangle$ = -1.09
and $\langle r^2\rangle_{2n}^{\rm so}$ = 0.062 fm$^2$.
Inserting this value into Eq. (\ref{chreq}), together with the 
core-dineutron distance obtained from the Coulomb dissociation 
experiment (last line of Table II), we now obtain the corrected 
value $\delta\langle r_{ch}^2\rangle_{\rm CD}$ = 0.88(11) fm$^2$
for the difference between the mean square charge radius of $^{11}$Li 
and $^9$Li.  This implies that the charge radius of $^{11}$Li
extracted from the Coulomb dissociation experiment is 2.41(4) fm,
which is consistent with the directly measured value of 2.467(37) fm
\cite{sanchez}.

The main part of the uncertainty in the charge radius of $^{11}$Li 
stems from the uncertainty in the charge radius of $^9$Li.
A better representation of the discrepancy we obtain in our three-body-model 
interpretation of the measured charge radius and dipole response of $^{11}$Li
is the difference
$$
\delta\langle r_{ch}^2\rangle_{\rm exp} -
\delta\langle r_{ch}^2\rangle_{\rm CD} 
=
$$
\begin{equation}
 1.10(8) - 0.88(11)  = 0.22(14) \ {\rm fm}^2.
\label{discr}
\end{equation}
This is the result we obtain when  we adopt revised finite mass 
corrections of Ref. \cite{moro}. The discrepancy is now of the 
order of $1.5\sigma$.

\section{Matter radius}

Also quoted in Table II is the calculated mean square distance between 
the two halo neutrons, $\langle r_{n,n}^2\rangle$, in $^{11}$Li. 
This quantity is not probed by the two experiments discussed above.
It is seen that this distance increases dramatically when the 
magnitude of the ground state s-wave component increases.

The RMS matter radius, obtained from an analysis of interaction
cross sections provides an additional constraint on the 
size of the halo. The mean square matter radius of a two-neutron 
halo nucleus is determined by the size of the halo and the core 
nucleus as follows
$$
\langle r_{m}^2(Z,A) \rangle = 
\frac{A-2}{A} \ \langle r_{m}^2(Z,A-2) \rangle 
$$
\begin{equation}
+ \ \frac{2(A-2)}{A^2} 
\langle r_{c,2n}^2 \rangle \ 
+ \ \frac{1}{2A}
\langle r_{n,n}^2 \rangle.
\label{rmat}
\end{equation} 
The RMS radii obtained in Ref.  \cite{tani} are 2.43$\pm$0.02 and 
3.27$\pm$0.24 fm, respectively, for $^9$Li and $^{11}$Li.
From the halo distances given in Table II and the quoted matter radius 
of $^9$Li we obtain a $^{11}$Li RMS matter radius of 3.29 fm in the 
$s23$ model. 
Thus the $s23$ model agrees with the matter radius and also with the 
strength of the dipole response but the charge radius is too small 
compared with the measured value.
The $s50$ model, on the other hand, agrees with the charge radius 
measurement, but the matter radius (which is 3.66 fm) and the dipole 
strength are too large compared to experiments.

\begin{figure}
\includegraphics [width = 8.6 cm]{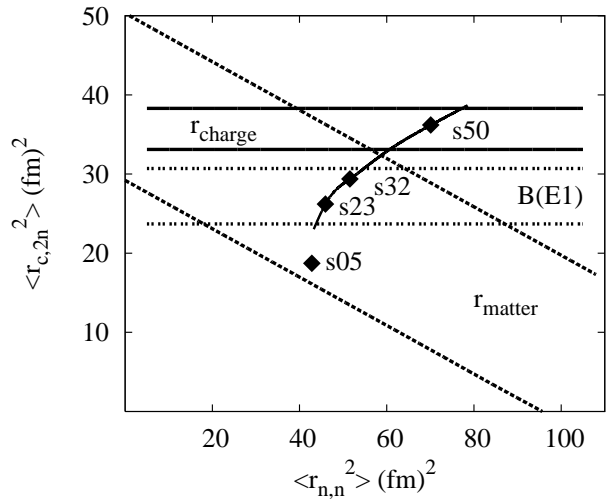}
\caption{Error bands on the size of the two-neutron halo in $^{11}$Li
obtained from the Coulomb dissociation experiment \cite{naka} (B(E1)),
the charge radius \cite{sanchez} ($r_{\rm charge}$),
and the matter radius \cite{tani} ($r_{\rm matter}$).
The results obtained in three-body models,
with 5 to 50\% s-wave components, are indicated by diamonds.}
\label{corli}
\end{figure} 

The constraints on the mean square distances of the two-neutron halo
obtained from the measurements of the charge radius, the dipole
response, and the matter radius of $^{11}$Li are illustrated in
Fig. \ref{corli} together with the predictions of the three-body
models. The limits from the charge radius are based on
the revised values \cite{moro} in Table II and do not
include the correction from the spin-orbit charge density.
That correction may reduce the charge radius limits on
$\langle r_{c,2n}^2\rangle$ by 1.9 fm$^2$, 
according to Eq. (\ref{chreq}), if we adopt the spin-orbit 
correction $\langle r^2\rangle_{2n}^{\rm so}$ = 0.062 fm$^2$,
which we obtained in the $s23$ model.

There is unfortunately some disagreement about the value of the 
matter radius which has been extracted from reaction data. An example 
is Ref. \cite{alkha} where RMS radii of 2.30 and 3.53$\pm$06 fm were 
obtained for $^9$Li and $^{11}$Li, respectively. 
This constraint provides a lower limit which is 
close to the upper limit of the matter radius shown in Fig. 3,
and it would therefore favor the $s50$ over the $s23$ model.
However, we do not think this is a reasonable solution because 
the $s50$ model produces a dipole response that has roughly a 33\%
larger strength than what has been observed 
(compare the values of $\langle r_{c,2n}^2 \rangle$ shown in Table II).

\section{Final remarks}

We think that the 1.5$\sigma$ discrepancy 
we obtain in our three-body model interpretation of the
measured charge radius and dipole response of $^{11}$Li 
is most likely caused by the neglect of core polarization.
Actually, it may seem surprising that the effect of core polarization 
is not much more dramatic. 

In order to estimate the effect of core polarization,
we have performed Skyrme Hartree-Fock calculations for $^9$Li and 
$^{11}$Li using the filling approximation and the SGII interaction.  
The results show that the mean-square charge radius of $^{11}$Li 
increases by 0.3 fm$^2$ from that of $^9$Li because of the core 
polarization effect, which is caused by the proton-neutron 
interaction when the valence neutrons occupy p-waves, whereas it 
increases by about 0.2 fm$^2$ for s-waves.  
This accounts roughly for the discrepancy, Eq. (\ref{discr}), 
between the charge radius and the Coulomb dissociation experiment.

We conclude that it would be desirable to extend the three-body
model so that one can consider the effect of core polarization 
in a consistent way. Work in this direction has already been done
for the ground state of $^{11}$Li \cite{kalman}, and it is also 
being pursued by other groups \cite{myo}.  
The work by Varga et al. \cite{kalman} shows that core polarization 
does play a significant role in their microscopic cluster model 
calculation of the charge radius of $^{11}$Li. 
This can be seen in Fig. 2 of Ref. \cite{sanchez}, where their results,
with and without the effect of core polarization, are compared to the 
measured charge radius.
A further test of such models is provided by the measured quadrupole 
moments of $^9$Li and $^{11}$Li \cite{neugart}, and by the dipole 
response that was extracted from Coulomb dissociation data \cite{naka}. 
In this connection, it would also be very useful to test the consistency 
of the measured charge radius and dipole response of $^6$He because the 
effect of core polarization should be much smaller for an $\alpha$ core.

\begin{acknowledgments} 
We are grateful to 
T. Nakamura for including the experimental energy resolution
in the calculated dipole response,
and to S. Shlomo and S. C. Pieper for reminding us 
of the significance of the spin-orbit charge density.
This work was supported (H.E. and P.M.) by the U.S. Department of Energy, 
Office of Nuclear Physics, under Contract No. DE-AC02-06CH11357,
and (K.H.) by the Grant-in-Aid for Scientific Research, 
Contract No. 19740115 from the Japanese Ministry of Education,
Culture, Sports and Technology.
\end{acknowledgments}


\begin{thebibliography}{99}
\bibitem{sanchez} R. Sanchez {\it et al}., Phys. Rev. Lett. {\bf 96}, 033002 (2006).
\bibitem{naka} T. Nakamura {\it et al}., Phys. Rev. Lett. {\bf 96}, 252502 (2006).
\bibitem{ann91} G. F. Bertsch and H. Esbensen, Ann. Phys. {\bf 209}, 327 (1991).
\bibitem{zhukov} M. V. Zhukov {\it et al}., Phys. Lett. B {\bf 265}, 19 (1991).
\bibitem{thompson} I. J. Thompson and M. V. Zhukov, Phys. Rev. C {\bf 49}, 1904 (1994).
\bibitem{young1} B. M. Young {\it et al}., Phys. Rev. Lett. {\bf 71}, 4124 (1993).
\bibitem{young2} B. M. Young {\it et al}., Phys. Rev. C {\bf 49}, 279 (1994).
\bibitem{neugart} D. Borremans et al., Phys. Rev. C {\bf 72}, 044309 (2005).
\bibitem{otsuka} T. Suzuki and T. Otsuka, Phys. Rev. C {\bf 56}, 847 (1997).
\bibitem{prc57} G. F. Bertsch, K. Hencken and H. Esbensen, Phys. Rev. C {\bf 57}, 1366 (1998).
\bibitem{simon} H. Simon {\it et al}., Phys. Rev. Lett. {\bf 83}, 496 (1999).
\bibitem{friar} J. L. Friar and J. W. Negele, Adv. Nucl. Phys. {\bf 8}, 219 (1975). 
\bibitem{PLB592} Particle Data Group, S. Eidelman {\it et al.}, Phys. Lett. B {\bf 592}, 1 (2004).
\bibitem{prc56} H. Esbensen, G. F. Bertsch and K. Hencken, Phys. Rev. C {\bf 56}, 3054 (1997).
\bibitem{wang} L.-B. Wang {\it et al.}, Phys. Rev. Lett. {\bf 93}, 142501 (2004).
\bibitem{ewald} G. Ewald {\it et al.}, Phys. Rev. Lett. {\bf 93}, 113002 (2004).

\bibitem{cobis} A. Cobis, D. V. Fedorov, and A. S. Jensen, 
Phys. Rev. Lett. {\bf 79}, 2411 (1997).
\bibitem{pieper} S. C. Pieper (private communication).
\bibitem{pudle} B. S. Pudliner, V. R. Pandharipande, J. Carlson, and 
R. B. Wiringa, Phys. Rev. Lett. {\bf 74}, 4396 (1995).
\bibitem{npa542} H. Esbensen and G. F. Bertsch, Nucl. Phys. {\bf 542}, 310 (1992).
\bibitem{prc46} H. Esbensen and G. F. Bertsch, Phys. Rev. C {\bf 46}, 1552 (1992).
\bibitem{hagprc72} K. Hagino and H. Sagawa, Phys.  Rev. C {\bf 72}, 044321 (2005). 
\bibitem{moro} M. Puchalski, A. M. Moro, and K. Pachucki, Phys. Rev. Lett.
{\bf 97}, 133001 (2006).
\bibitem{tani} I. Tanihata {\it et al}., Phys. Rev. Lett. {\bf 55}, 2676 (1985).
\bibitem{alkha} J. A. Tostevin and J. S. Al-Khalili, Nucl. Phys. A {\bf 616}, 
418c (1997).
\bibitem{kalman} K. Varga, Y. Suzuki, and R. G. Lovas, Phys. Rev. C {\bf 66},
041302(R) (2002).
\bibitem{myo} T. Myo, K. Kato, H. Toki, and K. Ikeda, Mod. Phys. Lett. A{\bf 21}, 2491 (2006).
\end{thebibliography}
\end{document}